\documentclass[dvipdfmx]{pasj01}

\Received{$\langle$13-Sep-2016$\rangle$}
\Accepted{$\langle$09-Jan-2017$\rangle$}
\Published{$\langle$dd-Mmm-yyyy$\rangle$}

\usepackage{color}
\usepackage{ulem}
\usepackage{threeparttable}

\newcommand{\Msun}{M_{\odot}}

\begin{document}
\SetRunningHead{Kumamoto et al.}{Interpretation of AVR}
\title{
Imprints of Zero-Age Velocity Dispersions and Dynamical Heating on the Age-Velocity dispersion Relation
}

\author{Jun \textsc{Kumamoto}$^{1,2}$,
Junichi \textsc{Baba}$^{2,3}$, and Takayuki R. \textsc{Saitoh}$^{2}$}
\affil{$^1$ Astronomical Institute, Tohoku University, Aramaki, Aoba, Sendai, Miyagi 980-8578, Japan.}
\affil{$^2$ Earth-Life Science Institute, Tokyo Institute of Technology, Ookayama, Meguro, Tokyo 152--8551, Japan.}
\affil{$^3$ Research Center for Space and Cosmic Evolution, Ehime University, Bunkyo-cho, Matsuyama, Ehime, 790--8577, Japan.}
\email{j.kumamoto@astr.tohoku.ac.jp}


\maketitle

\begin{abstract}
Observations of stars in the the solar vicinity show a clear tendency for old stars to
have larger velocity dispersions. 
This relation is called the age-velocity dispersion relation (AVR) and it is believed to 
provide insight into the heating history of the Milky Way galaxy.  
Here, in order to investigate the origin of the AVR, 
we performed smoothed particle hydrodynamic simulations of the self-gravitating multiphase gas disks
in the static disk-halo potentials. 
Star formation from cold and dense gas is taken into account, and we analyze the evolution of these star particles.
We find that exponents of simulated AVR 
and the ratio of the radial to vertical velocity dispersion are close to
the observed values.
We also find that the simulated AVR is not a simple consequence of dynamical heating. 
The evolution tracks of stars with different epochs evolve gradually in the age-velocity dispersion 
plane as a result of:
(1) the decrease in velocity dispersion in star forming regions, and
(2) the decrease in the number of cold/dense/gas as scattering sources.
These results suggest that the AVR involves not only the heating history of a stellar disk,
but also the historical evolution of the ISM in a galaxy.
\end{abstract}

\section{Introduction} \label{sec:Introduction}

It is well known that the velocity dispersion of stars in the solar vicinity
increases with stellar age. 
This is called the age--velocity dispersion relation (hereafter AVR). 
\citet{Stromberg1946} and \citet{SpitzerSchwarzschild1951} 
were the first to investigate the association between stellar velocity dispersion and spectral types, 
and then they discovered this relation.
Subsequent observations \citep{Nordstrom+2004, SeabrokeGilmore2007, Holmberg+2007, Holmberg+2009, AumerBinney2009,
JustJahreiss2010, Sharma+2014} confirmed the existence of the AVR with much larger samples.
According to these observations, this relation is well fitted by a single power-law function
for both the radial ($\sigma_R$) and vertical ($\sigma_z$) directions:
\begin{equation}
\sigma \propto \tau^\beta,
\end{equation}
where $\sigma$ represents $\sigma_R$ or $\sigma_z$ and $\tau$ is the age of
the star. The power-law index, $\beta$, is $0.3$--$0.5$ with certain errors
(See table 8 of \cite{Sharma+2014}).
The index for the radial direction is slightly smaller than that for the
vertical direction. However, it is not clear whether this difference is real or not.
$\sigma_z/\sigma_R \sim 0.5$ has also been ascertained by observations 
\citep{DehnenBinney1998}. 

Some secular heating processes and/or time evolution in the disk play crucial
roles in establishing the AVR 
(e.g., \cite{Wielen1977, FreemanBland-Hawthorn2002,Bland-HawthornGerhard2016} for reviews), 
since the older stars tend to have larger velocity dispersions. 
There are a number of scenarios for the origin of the AVR:  
heating by giant molecular clouds (GMCs; \cite{SpitzerSchwarzschild1951,
SpitzerSchwarzschild1953,KokuboIda1992}), by transient spiral arms \citep{CarlbergSellwood1985, DeSimone+2004},
by the combination of GMCs and spiral arms \citep{Carlberg1987, JenkinsBinney1990}, 
by halo black holes \citep{LaceyOstriker1985, HanninenFlynn2002},
or by minor mergers \citep{TothOstriker1992,Walker+1996, HuangCarlberg1997}. 
Until now, there has been no consensus about the primary source of the AVR. 
In table \ref{tab:AVR:Theory}, we summarize the theoretical predictions of 
velocity dispersions in both the radial and vertical directions.
All of the above processes are key features of galaxy formation.
Thus, understanding the origin of the AVR is essential in order 
to advance our understanding of galaxy formation.

\begin{table*}[htb]
  \caption{
  Velocity dispersions predicted by the theoretical models. 
  We summarize the models referring to AVRs and its exponents.
  Note that, in the case of minor mergers, only the amount of the increase of velocity dispersions 
  are shown. 
  }
  \begin{center}
  \begin{tabular}{lll} 
    \hline
                                           & heating source    & velocity dispersion $\sigma$\\ 
    \hline\hline 
    \cite{SpitzerSchwarzschild1953} & GMC               & $\sigma_{\rm mean} \propto \tau^{1/3}$                                          \\
    \cite{KokuboIda1992}            & GMC               & $\sigma_R \propto \tau^{0.25}$, $\sigma_z \propto \tau^{0.25}$              \\
    \cite{HanninenFlynn2002}        & GMC               & $\sigma_R \propto \tau^{0.21}$, $\sigma_z \propto \tau^{0.26}$              \\
    \cite{DeSimone+2004}            & spiral arm        & $\sigma_R \propto \tau^{0.20 {\rm \mathchar`-} 0.60}$                                      \\
                                           &                   & (depending on spiral properties)                                            \\
    \cite{JenkinsBinney1990}        & GMC \& spiral arm & $\sigma_R \propto \tau^{0.5}$, $\sigma_z \propto \tau^{0.3}$                \\
    \cite{HanninenFlynn2002}        & halo black hole   & $\sigma_R \propto \tau^{0.50}$, $\sigma_z \propto \tau^{0.50}$              \\
    \cite{Walker+1996}              & minor merger      & incease by $\delta \sigma \sim (10,8,8)~{\rm km s^{-1}}$ at solar radius    \\
    \cite{HuangCarlberg1997}        & minor merger      & $\sigma_z$ increase by a few -- dozens \%                                   \\
                                           &                   & (depending on satellite mass and falling angle)                             \\
    \hline
  \end{tabular}
  \end{center}
  \label{tab:AVR:Theory}
\end{table*}

Here, we mainly focus on the contribution of GMCs to the formation of the AVR.  
Some previous studies investigated the effect of gravitational scattering by GMCs 
on the velocity dispersions of stars.  
\citet{KokuboIda1992} calculated stellar orbits under the
influence of the gravitational potential of GMCs, and they found that the evolution
of velocity dispersion could be divided into two phases.  
In the early phase, the
evolution of the velocity dispersion is proportional to $\exp(\tau)$, while in
the late phase, it is proportional to $\tau^{0.25}$.
\citet{HanninenFlynn2002} performed $N$-body simulations of the stellar disk
involving GMCs modeled by spherical mass distributions of uniform density.  They
found that the radial and vertical velocity dispersions are proportional to
$\tau^{0.21}$ and $\tau^{0.26}$,
respectively. This is consistent with the prediction
from the simplified model of \citet{KokuboIda1992}, but the expected values of
velocity dispersions are somehow different from those obtained from observations.
Hence, gravitational scattering by GMCs is not considered
a primary source of the AVR.
However, the models used above show significant room for improvement.
For instance, they did not take into account gas dynamics, they did not model GMCs as realistic structures, 
and they did not consider star-formation processes.
\citet{Aumer+2016a} and \citet{Aumer+2016b} performed controlled $N$-body simulations of disk
galaxies to investigate the disk growth. 
In these simulations, they study the contribution of GMCs to the disk growth by using massive $N$-body 
particles which represent GMCs.
They found that the efficiency of GMC heating depends on the fraction of disk mass residing in the GMC.
Models that are more realistic are necessary 
to conclude whether or not 
GMC heating is important in the establishment of the AVR.

Recently, some zoom-in cosmological simulations
have been used to derive numerical estimations of the AVRs.
\citet{Brook+2012} carried out $N$-body/smoothed particle hydrodynamics (SPH)
simulations with the mass and spatial resolutions of $7300~\Msun$ and $155~{\rm pc}$.
They showed that old stars that have larger velocity dispersion were born at lower radii
and are kinematically hotter than those born at a later time.
\citet{Martig+2014} used adaptive mesh refinement code RAMSES \citep{Teyssier2002}
and they showed that AVR is not determined by the stellar velocity dispersion at birth time,
but is the result of subsequent heating.
Furthermore, \citet{House+2011} investigated the variations of AVR using
different modeling techniques. 
\citet{Grand+2016} used the state-of-the-art simulation code, {\tt AREPO} \citep{Springel2010}, for
their zoom-in simulations and reported that they obtained an AVR that is
consistent with the observed AVR.
While all studies showed power-law-like AVRs, we note that their simulations are
insufficient to disclose the origin of the AVR because of the limited numerical
resolutions (the force resolution is $\ge 100~{\rm pc}$), which is larger than 
the thickness of the molecular gas disk ($48$--$160~{\rm pc}$) in the Milky Way galaxy
\citep{NakanishiSofue2006}. 
Some models did not directly solve the low temperature ($\lesssim 100$ K) 
and high-density regions ($\gtrsim 100~\rm cm^{-3}$) that mimic GMCs.
In addition, the origin of the AVR is not well investigated in these studies.

In this paper, we find out the origin of the AVR by using high-resolution
$N$-body/SPH simulations. Instead of doing the cosmological simulations, 
we adopt an idealized model in which only gas and stars formed from cold and dense
gas are solved. The halo and (the main body of) the stellar disk are expressed by
static potentials. This approximation allows us to use high mass and
spacial resolutions, such as $6000~\Msun$ and $10~{\rm pc}$. 
As a trade-off, we cannot include the contribution of the spiral arms at present,
but this will constitute the next step of this study and 
we will investigate it in a forthcoming paper.

The structure of this paper is as follows. 
In \S \ref{sec:Model}, we describe the initial conditions of our simulations and our numerical methods. 
We present the evolution of the simulated galactic disk and AVRs in our simulation in \S \ref{sec:Result}.  
Heating rates are also discussed in this section. 
Finally, in \S \ref{sec:Summary}, we show the summary of our results and its
implications for the history of the formation and evolution of the Milky Way galaxy.

\section{Models and Methods}\label{sec:Model}

Our three dimensional (3D) $N$-body/SPH
simulations conducted in this study were performed using the {\tt ASURA-2}
\citep{SaitohMakino2009,SaitohMakino2010}.  
We investigated the AVR for stellar particles formed 
from SPH particles moving within a static halo and disk potential.

\subsection{Galaxy Model}

In order to concentrate on the contribution of the clumpy interstellar medium (ISM)
to the stellar velocity dispersion, 
we solve the evolution of the gas with high resolutions. 
For this purpose, here we assume that the halo and the main body of
the stellar disk are described by fixed potentials. 
This modeling is identical to that of \citet{Saitoh+2008}.

We assume that the dark matter halo follows the Navarro-Frenk-White (NFW)
profile \citep{Navarro+1997}, which is expressed as
\begin{eqnarray}
\rho_{\rm halo}(x)&=&\frac{\rho_{\rm c}}{x(1+x)^2}, \\ \label{eq:halo}
  x&=&\frac{r}{r_{\rm s}},
\end{eqnarray}
where $\rho_{\rm c}$ and $r_{\rm s}$ are the characteristic density and a scale radius of
this profile, respectively.  We assume that $\rho_c = 4.87\times10^6~\Msun~{\rm
kpc^{-3}}$ and $r_{\rm s}=21.5~{\rm kpc}$.

The potential of the stellar disk is described by the Miyamoto-Nagai model \citep{MiyamotoNagai1975}:
\begin{eqnarray}
  \rho_{\ast}(R,z)&=&\frac{M_{\ast}z_{\ast}^2}{4\pi}
  \times  \nonumber \\
 &&\frac{R_{\ast}R^2+(R_{\ast}+3\sqrt{z^2+z_{\ast}^2})(R_{\ast}+\sqrt{z^2+z_{\ast}^2})^2}
{[R^2+(R_{\ast}+\sqrt{z^2+z_{\ast}^2})^2]^{5/2}(z^2+z_{\ast}^2)^{3/2}},
  \label{eq:disk}
\end{eqnarray}
where $M_{\ast}$, $R_{\ast}$, and $z_{\ast}$ are the stellar disk mass, 
the scale radius and the scale height, respectively. We use $M_{\ast} = 4 \times
10^{10}~\Msun$, $R_{\ast} = 3.5~{\rm kpc}$, and $z_{\ast} = 400~{\rm pc}$.

\begin{figure}[htb]
\begin{center}
\includegraphics[width=0.48\textwidth]{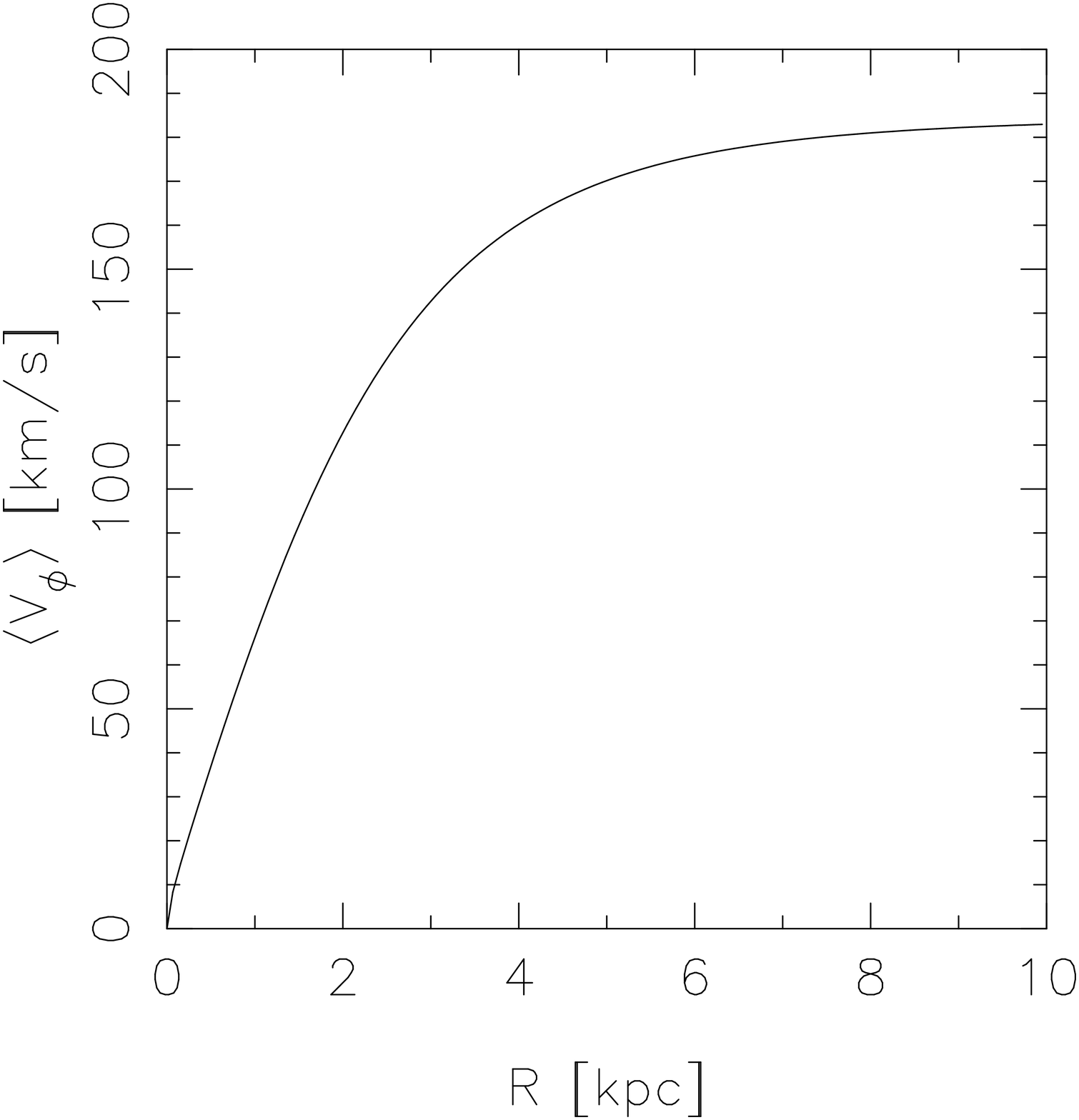}
\end{center}
\caption{
Initial rotational velocity as a function of radius.
$\langle v_{\phi}\rangle$ is the average azimuthal velocity of initial SPH particles 
located in each radial bin. We apply a width of $0.1~{\rm kpc}$ to the radial bins.
}
\label{fig:rotation}
\end{figure}

Initially, the gas disk has simple exponential and
Gaussian distributions in the radial and vertical directions.
The scale radius
and the scale height of the gas distribution are set to $7~{\rm kpc}$ and
$400~{\rm pc}$, respectively. 
Figure \ref{fig:rotation} shows the averaged rotation curve of gas particles at the initial conditions. 
The total mass of the gas disk is $6\times10^9\Msun$, 
and we use 1 million SPH particles to express this gas disk. 
Hence, the mass of each SPH particle is $6000~\Msun$.  
The gravitational softening length is set to $10~{\rm pc}$.

\subsection{Methods}

We solve the evolution of gas and star particles in the halo and disk potential
using our $N$-body/SPH code, 
{\tt ASURA-2} \citep{SaitohMakino2009,SaitohMakino2010}.  
The numerical technique is almost the same as those used in above papers, 
but with slight updates. We now briefly describe our method.

The self-gravity among the gas and star particles is calculated by the tree
with the GRAvity PipE (GRAPE) method \citep{Makino1991}. The opening angle is
set to 0.5 and only monopole moment is considered.
In order to accelerate the computation, 
we adopt the Phantom-GRAPE library \citep{Tanikawa+2013}, 
which is a software emulator of GRAPE.

The gas dynamics are solved by using SPH \citep{Lucy1977,GingoldMonaghan1977,Monaghan1992}.  
The cubic-spline kernel function is utilized.  
For the first space derivative of the kernel, Thomas \& Couchman's modification is employed
\citep{ThomasCouchman1992}.  The number of neighboring particles for each SPH
particle is kept within $32\pm2$ in the support radius of the kernel.  
The shock is handled with an artificial viscosity term proposed by \citet{Monaghan1997}. 
The viscosity coefficient is set to 1.0.
We also introduce the Balsara limiter in order to reduce the unwanted angular
momentum transfer \citep{Balsara1995}.

The time integration is conducted
by a second order scheme (see appendix A in \cite{SaitohMakino2016}). 
For SPH particles, we use the FAST scheme so that we can accelerate 
the time-integration in supernova (SN)-heated regions 
by using different time-steps for gravitational and hydrodynamical interactions
\citep{SaitohMakino2010} and the time-step limiter so that we can follow the
shocked region correctly \citep{SaitohMakino2009}.

In simulations in this paper, we adopt the radiative cooling, star formation,
and type II supernovae (SNe) feedback.  The radiative cooling of the gas is
solved by assuming an optically thin cooling function, $\Lambda(T,f_{\rm H2},G_0)$, 
for a wide temperature range of $20~{\rm K} < T < 10^8~{\rm K}$ \citep{Wada+2009}
For simplicity, we fix the molecular hydrogen fraction at $f_{\rm H2} = 0.5$, 
and the far-ultraviolet intensity is normalized to the solar neighborhood $G_0 = 1$.
We do not have to use the artificial pressure floor method, which is highlighted 
in \citet{Saitoh+2006} and \citet{RobertsonKravtsov2008}.
\footnote{The idea to introduce the pressure floor to simulations might come from the 
numerical experiments shown by \citet{BateBurkert1997}. They pointed out for the first time that 
artificial fragmentation might occur, if SPH simulations did not have sufficient mass resolution. 
After this indication, the pressure floor is introduced, in particular to avoid artificial fragmentations. 
However, \citet{Hubber+2006} conclude that under the insufficient mass resolution case, 
the growth of gravitational instability is suppressed when the artificial fragmentation did not take place.
}
Star formation is modeled in a probabilistic manner
following the Schmidt law (e.g., \cite{Katz1992,Saitoh+2008}).  An SPH particle
which satisfies the following three conditions, (1) $n_{\rm H} > 100~{\rm cm^{-3}}$; 
(2) $T < 100~{\rm K}$; and (3) $\nabla\cdot\mathbf{v}<0$, 
spawns a star particle with the Salpeter initial mass function \citep{Salpeter1955} and
is within the mass range of $0.1~\Msun < m_{\rm \star} < 100~\Msun$.  
As an early stellar feedback, the ${\rm H_{II}}$-region feedback 
using a Stromgren volume approach \citep{Baba+2017} is employed.  
The energy feedback from type II SNe is implemented. 
Each SN releases $10^{51}~{\rm erg}$, and this energy is injected
into the surrounding ISM.  We adopt the probabilistic manner of
\citet{Okamoto+2008} to evaluate the injection time.

\section{Results} \label{sec:Result}

We analyze the AVR in the simulations and investigate its evolution in \S \ref{sec:AVR}.
We then discuss the origins of AVR evolution in \S \ref{sec:AVROrigin}.

\begin{figure*}[htb]
\begin{center}
\includegraphics[width=0.96\textwidth]{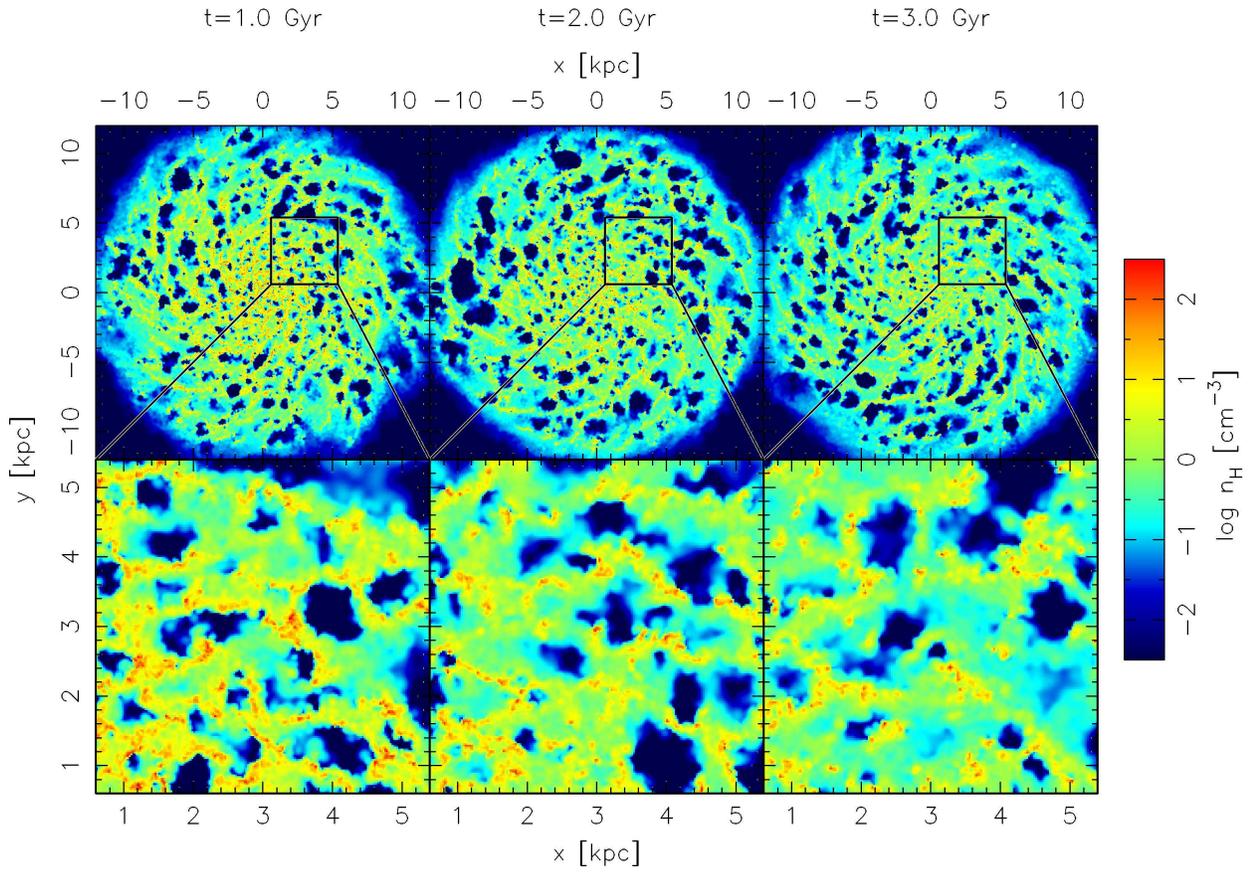}
\end{center}
\caption{
The gas density on the galactic plane ($z=0$) at $t=1$, $2$, and $3~{\rm Gyr}$.  
The upper panels show a region of $-12~{\rm kpc} < x < 12~{\rm kpc}$ and $-12~{\rm kpc} < y < 12~{\rm kpc}$.  
The lower panels display the close up view of a part of the disk region of $0.6~{\rm kpc} < x < 5.4~{\rm kpc}$ and $0.6~{\rm kpc} < y < 5.4~{\rm kpc}$.
}
\label{fig:ssG}
\end{figure*}

Before presenting the details of the AVR and its origin, we first explain 
the global evolution of our galaxies. Figure \ref{fig:ssG} shows 
the evolution of our simulated galaxies.  
The upper and lower panels show overviews and close-up views of the gas density on the
galactic plane ($z=0$).
From the close-up views of the gas density, we can see rich inhomogeneous structures of the ISM.
These structures are induced by our modeling with high resolution \citep{Saitoh+2008}. 
From $t = 1$ -- $3~{\rm Gyr}$,
the contrast of the gas density decreases because of the gas consumption 
due to star formation.  
This implies that the primary source heating the disk stars
decreases with increasing time in our simulation.
We discuss the relation between the contrast of the gas density and heating rate in \S \ref{sec:HeatingRate}.

\begin{figure*}[htb]
\begin{center}
\includegraphics[width=0.96\textwidth]{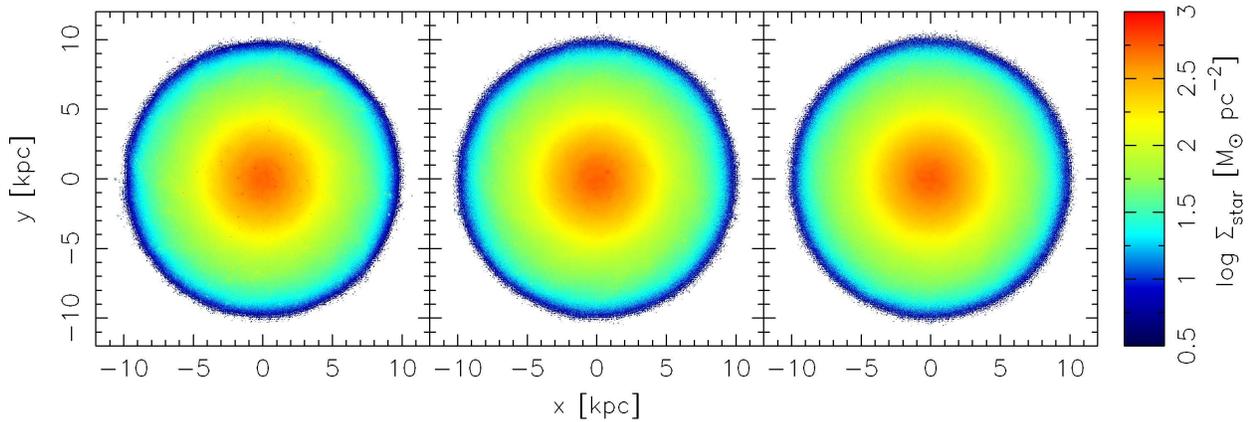}
\end{center}
\caption{
The stellar surface density at $t=1$, $2$, and $3~{\rm Gyr}$.
For the stellar surface density, we use the sum of stellar particles formed from the gas particles
and static surface density calculated by the Miyamoto-Nagai model \citep{MiyamotoNagai1975}. 
}
\label{fig:ssS}
\end{figure*}

The high-density regions exceed the star formation threshold density (i.e. $n_{\rm H} > 100~\rm cm^{-3}$)
Therefore, stars are formed in these high-density regions when the
other two star-formation conditions are satisfied.
Figure \ref{fig:ssS} shows the evolution of stellar surface density.
The stellar surface density distribution maps show no strong spiral arms developed by 
stellar dynamical mechanisms such as the swing-amplification \citep{Toomre1981} and 
non-linear interactions between wakelets \citep{KumamotoNoguchi2016}.
As the skeletal structure of the disk is constructed of the smoothed and 
axisymmetric Miyamoto-Nagai potential, it is difficult to enhance the spiral arms 
and a bar in this model; thus, it has a flocculent structure.

\begin{figure}[htb]
\begin{center}
\includegraphics[width=0.48\textwidth]{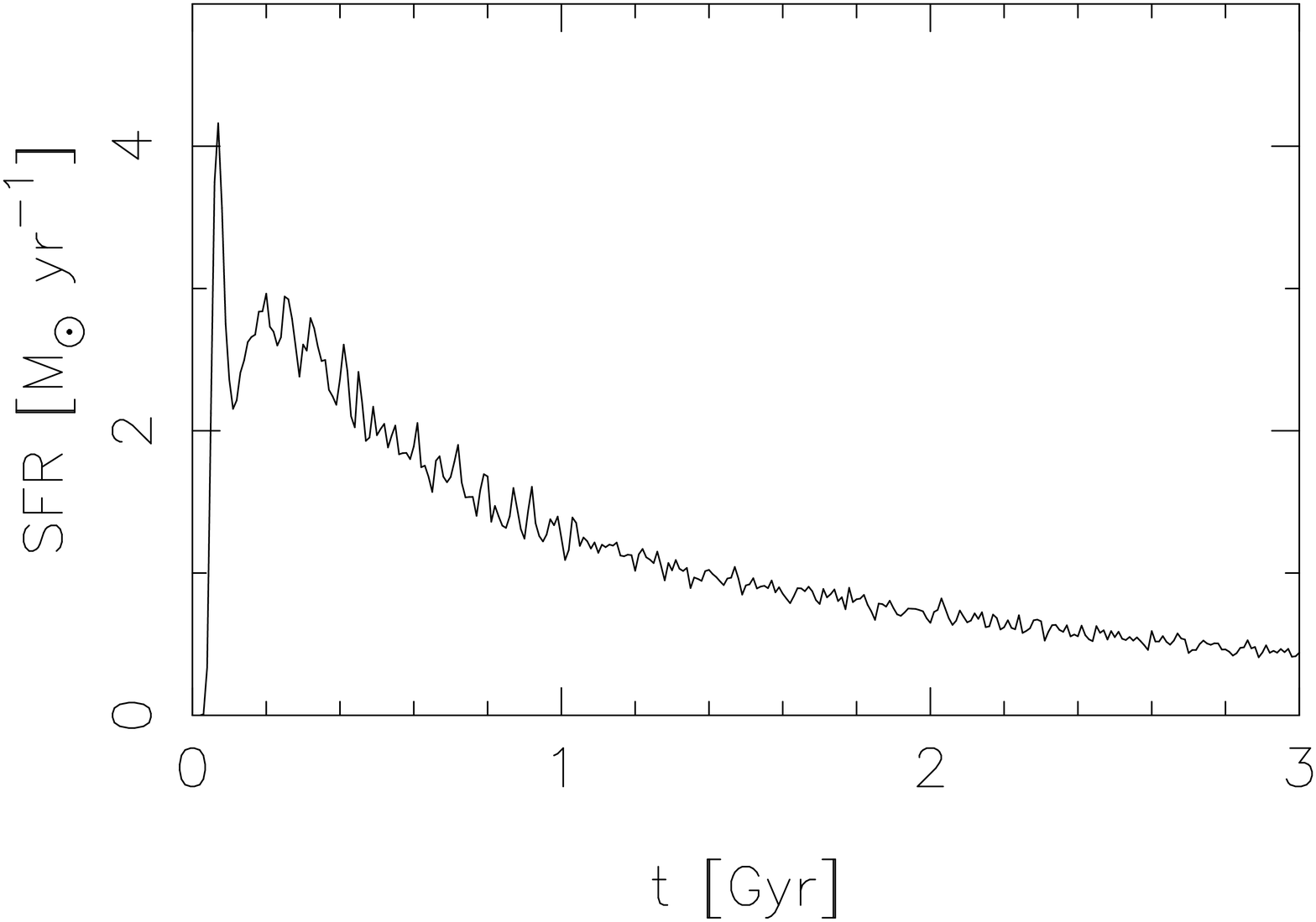}
\end{center}
\caption{
Star-formation rate as a function of time.
}
\label{fig:sfr}
\end{figure}

Figure \ref{fig:sfr} shows the star-formation rate as a function of time.  
It is characterized by two phases: the initial rapidly increasing phase ($t<100~{\rm Myr}$), 
and the succeeding exponentially decreasing phase.  
The decay time-scale of the star-formation rate is $\sim 1~{\rm Gyr}$. 
This time scale is rather short compared to the estimation of the galactic chemical evolution
\citep{Larson1972,Yoshii+1996}, as our simulation does not 
consider the secular evolution due to gas accretion from the halo.
This is also the reason why we stop our simulation at $t=3~{\rm Gyr}$, 
before the gas is exhausted. 
A model incorporating gas accretion will be used in our forthcoming papers.

\subsection{AVR}
\label{sec:AVR}

\begin{figure*}[htb]
\begin{center}
\includegraphics[width=0.96\textwidth]{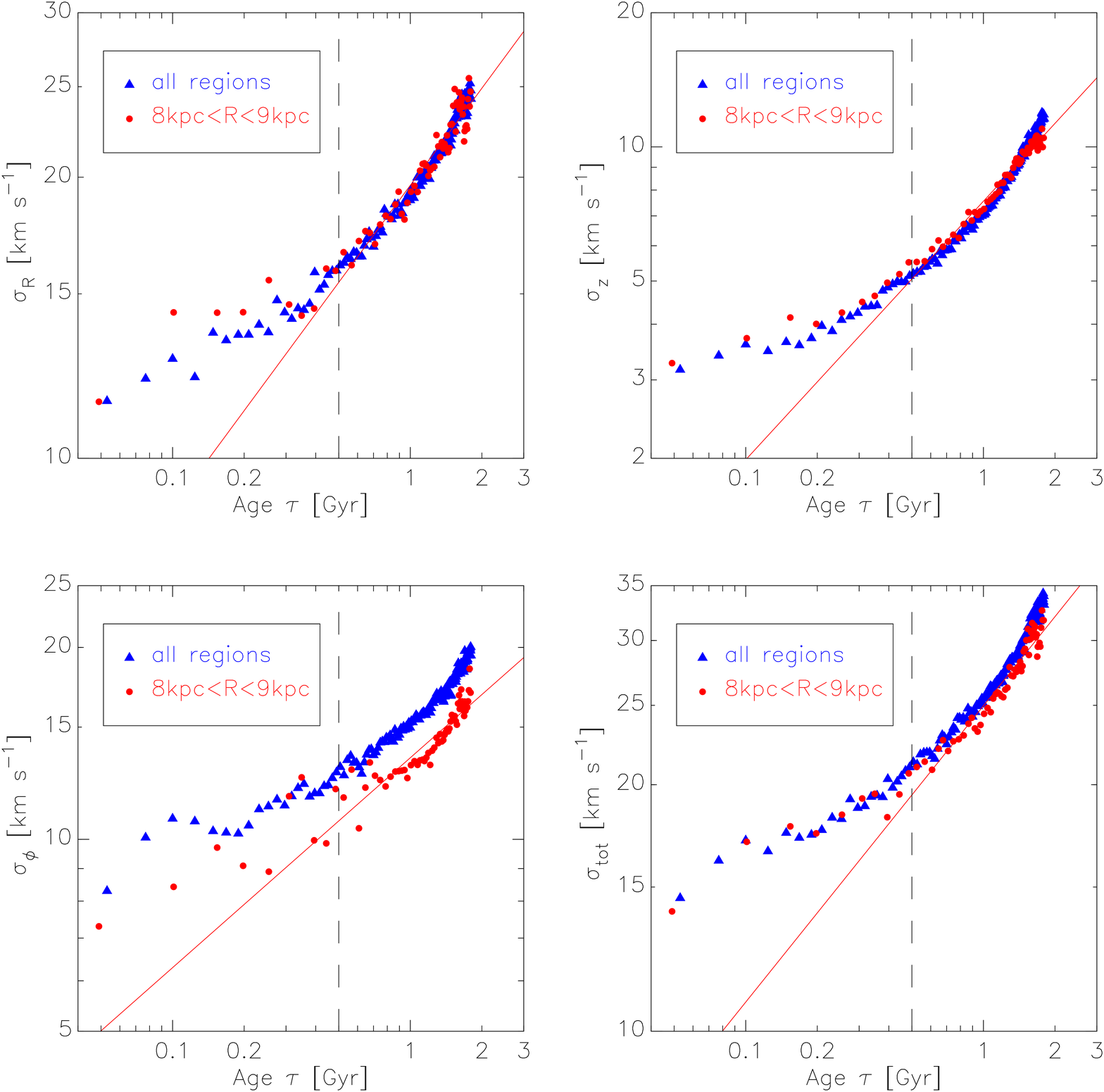}
\end{center}
\caption{
Radial (top left), vertical (top right), azimuthal (bottom left) and total (bottom right) 
AVRs in our simulated galaxy at $t = 2~\rm{Gyr}$. 
The red dots indicate the AVR calculated with stellar particles
found between $8~{\rm kpc}$ and $9~{\rm kpc}$ from the galactic center.
The solid lines are the fitting results with a power-law function 
for the stellar particles in this region.
For this procedure, we only use star particles older than $0.5~\rm{Gyr}$ 
(on the right side of the dashed line).
The blue triangles indicate the AVR calculated with all stellar particles. 
}
\label{fig:avr_all_ring}
\end{figure*}

Figure \ref{fig:avr_all_ring} shows the radial, azimuthal, vertical and total AVRs in our simulated
galaxy at $t = 2~\rm{Gyr}$.  In this AVR analysis, we only use the star particles formed from gas.
The red dots denote the results when only stars whose 
galactocentric distance $R$ was in the range $8~{\rm kpc} < R < 9~{\rm kpc}$ were used, 
whereas the blue triangles represent the case where all regions were included.
The velocity dispersions as a function of stellar age, $\tau$, are
calculated by using the following equations:
\begin{equation}
\sigma_{\rm R}(\tau) = \left[ \langle v_R^2 \rangle_{\tau_l < \tau < \tau_h} - 
	\langle v_R \rangle^2_{\tau_l < \tau < \tau_h} \right]^{1/2},
\end{equation}
\begin{equation}
\sigma_{\rm \phi}(\tau) = \left[ \langle (v_\phi- \overline{v_{\phi}}(R))^2 \rangle_{\tau_l < \tau < \tau_h} - 
	\langle v_\phi- \overline{v_{\phi}}(R) \rangle^2_{\tau_l < \tau < \tau_h} \right]^{1/2},
\end{equation}
\begin{equation}
\sigma_{\rm z}(\tau) = \left[ \langle v_z^2 \rangle_{\tau_l < \tau < \tau_h} - 
	\langle v_z \rangle^2_{\tau_l < \tau < \tau_h} \right]^{1/2},
\end{equation}
\begin{equation}
\sigma_{\rm tot}(\tau) = \left[ \sigma_{\rm R}(\tau)^2 + \sigma_{\rm \phi}(\tau)^2 + \sigma_{\rm z}(\tau)^2 \right]^{1/2},
\end{equation}
where $\tau_l$ and $\tau_h$ are the two edges of the age bin.
$\langle v_R^2 \rangle_{\tau_l < \tau < \tau_h}$ mean the average of stellar $v_R^2$ whose age is
between $\tau_l$ and $\tau_h$.
We apply the width of age bins in such a way that every bin has equal numbers of stars
(2000 stars in each bin).
For the distributions of velocity dispersions, the sampling range effect is insignificant.

The solid lines in figure \ref{fig:avr_all_ring} represent the fitting results with
a power-law function for stellar particles found between $R=8~{\rm kpc}$ and
 $9~{\rm kpc}$.  Here, we adopt stars older than $0.5~{\rm Gyr}$.  The functional
forms of the fitted age-radial, azimuthal, vertical, and total velocity dispersion relations are
\begin{equation}
  \sigma_R = 19.4 \left( \frac{\tau}{1~\rm{Gyr}} \right)^{0.37} ~{\rm km~s^{-1}},
  \label{eq:AVR:R}
\end{equation}
\begin{equation}
  \sigma_\phi = 13.4 \left( \frac{\tau}{1~\rm{Gyr}} \right)^{0.33} ~{\rm km~s^{-1}},
  \label{eq:AVR:phi}
\end{equation}
\begin{equation}
  \sigma_z = 7.36 \left( \frac{\tau}{1~\rm{Gyr}} \right)^{0.59}~{\rm km~s^{-1}},
  \label{eq:AVR:z}
\end{equation}
\begin{equation}
  \sigma_{\rm tot} = 25.0 \left( \frac{\tau}{1~\rm{Gyr}} \right)^{0.36} ~{\rm km~s^{-1}}.
  \label{eq:AVR:tot}
\end{equation}
The exponents of the radial, azimuthal, vertical, and total AVR 
and the ratio of the radial to vertical velocity dispersion are close to
the observed values (see \S \ref{sec:Introduction}). 
The radial to vertical velocity dispersion ratio becomes 
\begin{equation}
  \frac{\sigma_z}{\sigma_R} = 0.38 \left( \frac{\tau}{1~\rm{Gyr}} \right)^{0.22}.
  \label{eq:AVR:ratio}
\end{equation}
Observationally, the age dependence on this relation is not found.  This weak
dependence on $\tau$ in our simulated AVR again favors observations 
\citep{DehnenBinney1998}.

The blue triangles in figure \ref{fig:avr_all_ring} show the AVR
calculated with all stellar samples.
Here, we reapply the width of age bins in such a way that each bin has 10000 stars.
Radial and vertical AVRs are similar to the local AVRs shown by
the red points in spite of the difference in the samples.
This agreement means that radial and vertical velocity dispersions are 
roughly constant throughout the disk region.

\begin{figure}[htb]
\begin{center}
\includegraphics[width=0.48\textwidth]{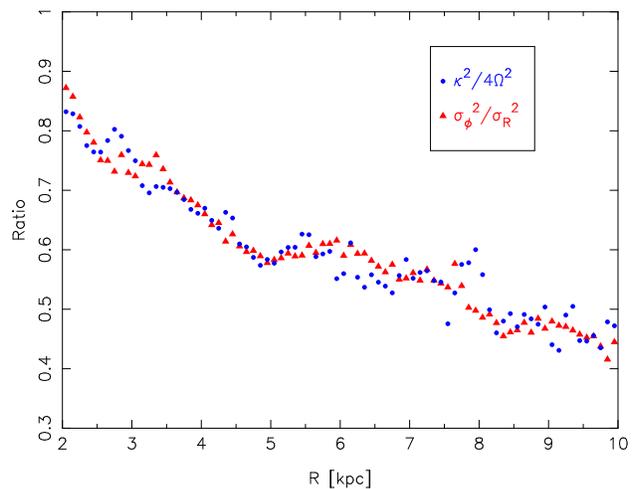}
\end{center}
\caption{
$\sigma_{\rm \phi}^2 / \sigma_{\rm R}^2$ (red triangles) and $\kappa^2 / 4\Omega^2$ (blue dots)
as a function of galactocentric radius at $t=2~{\rm Gyr}$.
These two parameters are equal when epicycle approximation is satisfied.
}
\label{fig:epicycle}
\end{figure}

On the other hand, azimuthal velocity dispersion with all stellar samples is larger than those with local stars 
in the outer region ($8~{\rm kpc} < R < 9~{\rm kpc}$). 
That is because the azimuthal velocity dispersion is determined by the epicycle approximation \citep{BinneyTremaine2008},
\begin{equation}
  \frac{\sigma_{\rm \phi}^2}{\sigma_{\rm R}^2} = \frac{\kappa^2}{4\Omega^2} ,
  \label{eq:epicycle}
\end{equation}
where $\kappa$ and $\Omega$ are epicycle frequency and angular velocity, respectively, 
and the ratio of $\kappa$ and $\Omega$ changes depending on the galactocentric distance.
Figure \ref{fig:epicycle} shows $\sigma_{\rm \phi}^{2}/\sigma_{\rm R}^2$ and $\kappa^2/\Omega^2$ as a function of radius.
There is a good agreement between then, which means that stellar motion is well described with 
the epicycle approximation. We see that $\sigma_{\rm \phi}^2/\sigma_{\rm R}^2 \sim 0.5$ in the outer 
region ($R>8~{\rm kpc}$) where the rotational velocity becomes almost flat (as shown in figure \ref{fig:rotation}), and this is also 
consistent with the prediction of the epicycle approximation.
In the inner region ($R<8~{\rm kpc}$), however, the rotational velocity is an increasing function of R,
and then $\sigma_{\rm \phi}^{2}/\sigma_{\rm R}^2$ are larger than those in the outer region.
Here, $\sigma_{\rm R}$ does not depend on radius as mentioned in the above paragraph, and so, 
$\sigma_{\rm \phi}$ becomes a decreasing function of radius according to the epicycle approximation.
Therefore, azimuthal AVR with all stellar samples deviates upward from
that with local stars between $8~{\rm kpc}$ and $9~{\rm kpc}$ as shown in figure \ref{fig:avr_all_ring}.

Hereafter, we discuss the radial and vertical AVR with all stellar particles in the galactic disk to
ensure sufficient samples and simplification.

It is worth noting that strong spiral arms do not develop in our simulated galaxies, 
although they are regarded as a heating source of radial velocity dispersion
\citep{Carlberg1987, JenkinsBinney1990}.  Interestingly, without strong spiral
arms, we can successfully reproduce the observed exponents of 
not only the vertical AVR, but also the radial one.  
We discuss this point in \S \ref{sec:Summary}.

\begin{figure*}[htb]
\begin{center}
\includegraphics[width=0.96\textwidth]{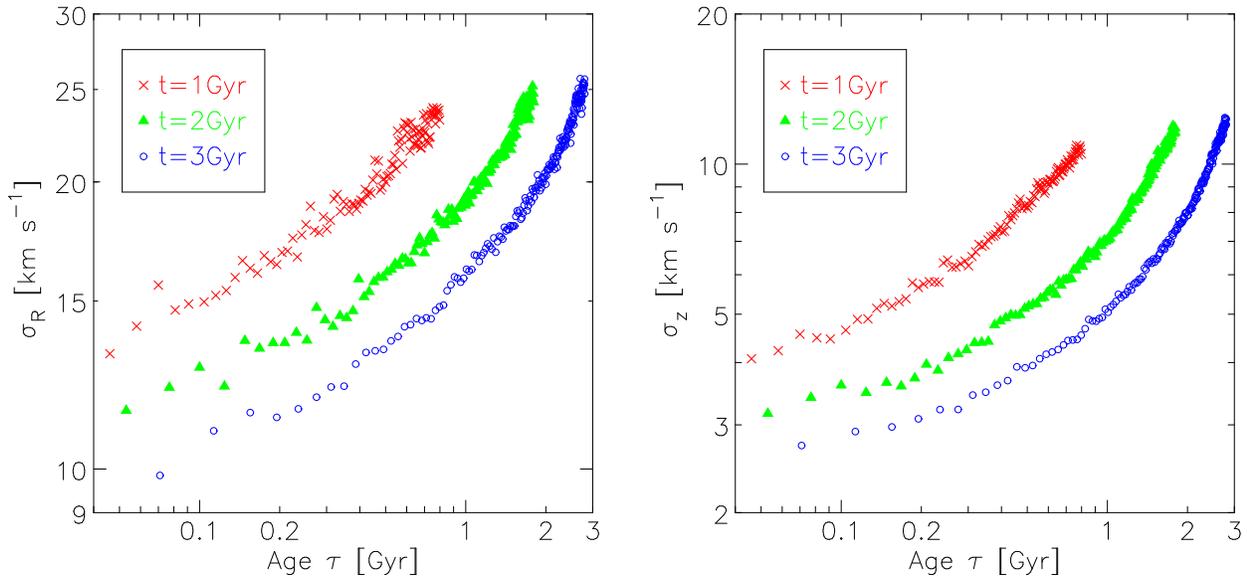}
\end{center}
\caption{
Evolution tracks of stellar age-velocity dispersions with different samples
whose birth epoch is $t=1$, $2$, and $3~\rm{Gyr}$.  Different symbols represent
different groups.
}
\label{fig:AVR:Evolution}
\end{figure*}

Figure \ref{fig:AVR:Evolution} shows the AVRs at the three different epochs.  
We can see that the simulated AVRs of the three epochs do not overlap; 
there is a clear indication of the evolution of the AVR. 
Both the initial velocity dispersions and heating efficiency evolve. 
This is inconsistent with the conventional interpretation of the AVR, i.e.,
that it is the evolution track of stars on the age--velocity dispersion
plane.

\begin{figure*}[htb]
\begin{center}
\includegraphics[width=0.96\textwidth]{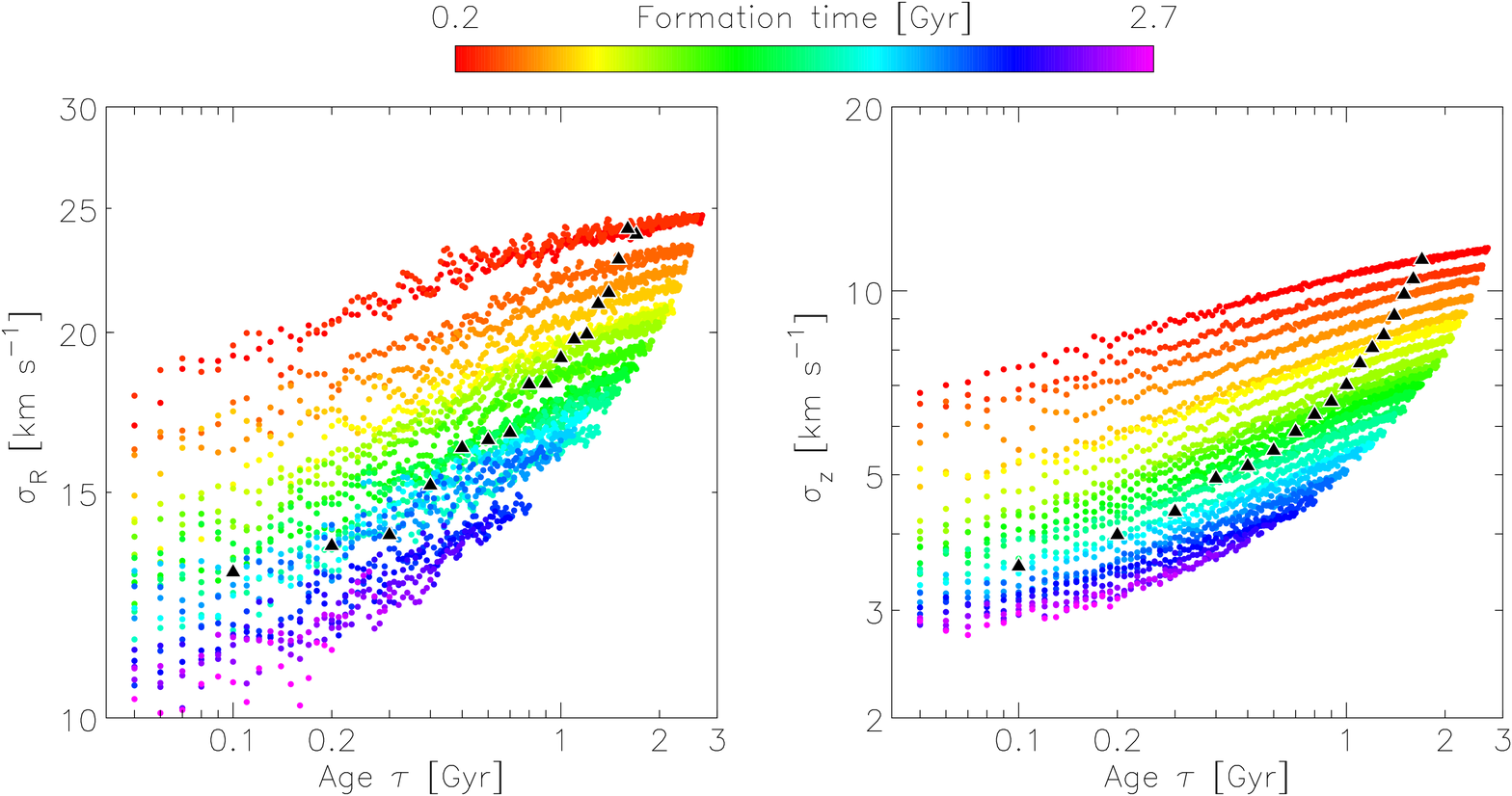}
\end{center}
\caption{
Dots show the time evolution of stellar velocity dispersion.  
The left and right panels show the radial and 
vertical velocity dispersion, respectively.  
Each color indicates a different birth time.  
The black triangle points show the AVR at $t=2$ Gyr.
}
\label{fig:AVR:Evolution2}
\end{figure*}

In figure \ref{fig:AVR:Evolution2}, we show the evolution tracks of stellar
groups of 26 different epochs on the age--velocity dispersion planes.  Each group
has a different initial velocity dispersion and heating rate.  
The slope of each evolution sequence is less steep 
than those in observations.  
However, when we pick up values, for instance, at $t=2~{\rm Gyr}$ for all evolution sequences 
(black triangles in figure \ref{fig:AVR:Evolution2}), 
we obtain AVRs that are consistent with those
obtained from observations. This reveals that AVRs are inconsistent with 
the evolution tracks of stars on the age-velocity dispersion plane.

\subsection{Origins of simulated AVR} \label{sec:AVROrigin}

In the previous section, we showed that the AVR is not just a simple evolutionary track 
of stars on the age--velocity dispersion plane. 
In this section, we investigate two important factors that affect 
the evolution of stars on the age--velocity dispersion plane:
the initial velocity dispersion (i.e. velocity dispersion at the birth time of stars)
and the heating rate as a function of age.  The
evolution of the initial velocity dispersion is discussed in \S \ref{sec:InitialVelocityDispersion}, 
while that of the heating rate is discussed in \S \ref{sec:HeatingRate}.

\subsubsection{Zero-Age Velocity Dispersion}
\label{sec:InitialVelocityDispersion}

First, we discuss the stellar velocity dispersion at the birth time of stars. 
We call this the ``zero-age velocity dispersion'' (hereafter, ZAVD).  
As shown in figure \ref{fig:AVR:Evolution2}, the ZAVD decreases with increasing age. 
This is one of the key reasons for the inconsistency between the AVR and the evolution
tracks.

\begin{figure}
\begin{center}
\includegraphics[width=0.48\textwidth]{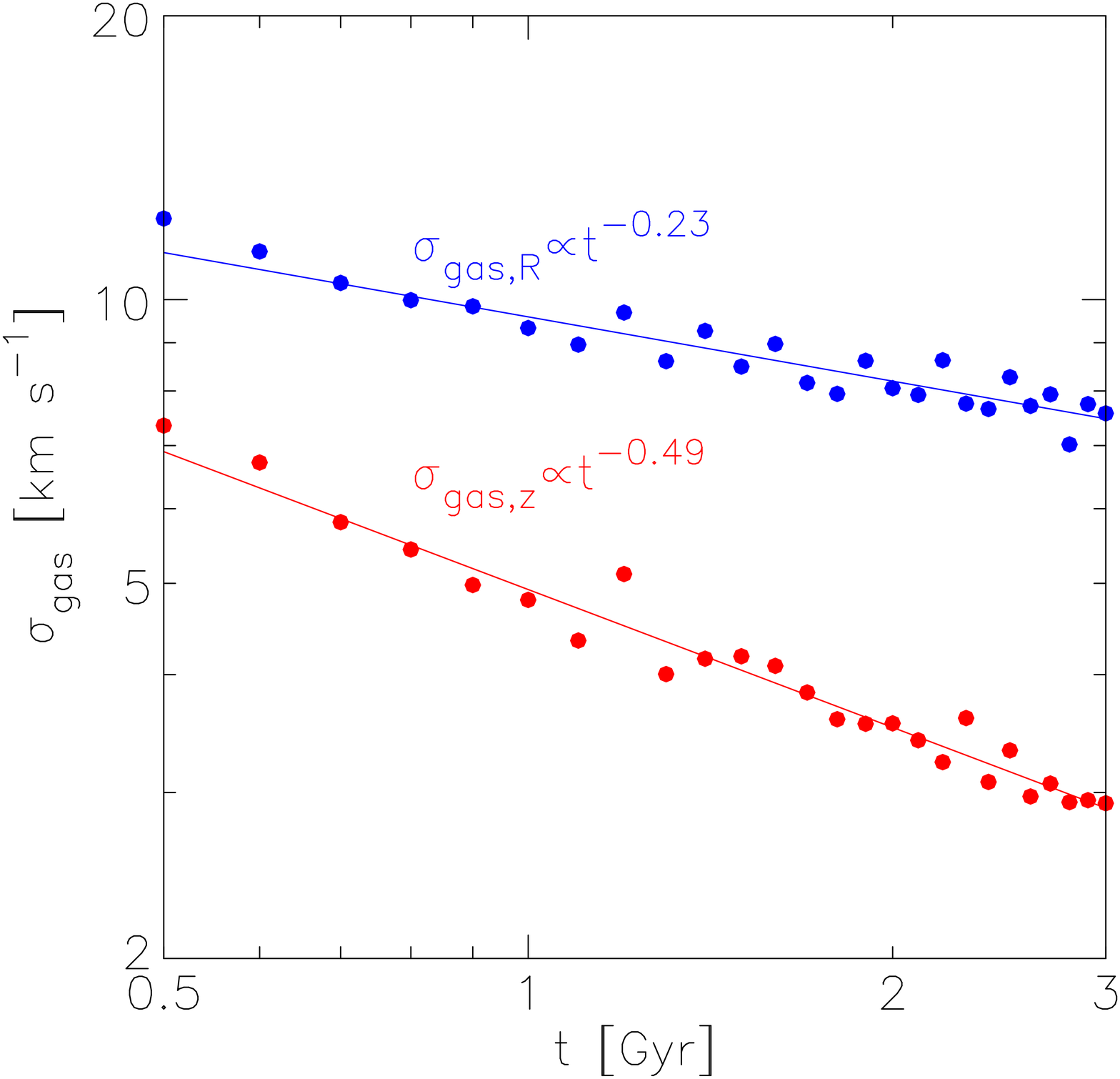}
\end{center}
\caption{
Velocity dispersion of gas that satisfies the star formation criteria as a
function of time. Both the radial and vertical velocity dispersions are shown.
Blue dots represent the radial-velocity dispersions,
while red dots indicate the vertical-velocity dispersions. 
The lines are the results of the least square
fitting with a power-law function.
}
\label{fig:gas_sigma}
\end{figure}

The ZAVD is strongly related to the velocity dispersion of 
the star-forming gas because it inherits the velocity dispersion. 
Figure \ref{fig:gas_sigma} shows the radial and vertical velocity dispersions of gas 
that satisfy the star-formation criteria as a function of time. 
We can see that both velocity dispersions monotonically decrease with time 
and they can be fitted by a power-law function of $t$.
However, their power-law indices are different; the index of the ZAVD for the
vertical direction is about twice as steep as that for the radial direction.
Gravitational interactions, radiative cooling, star formation and feedback play
crucial roles in determining the ISM structure, and these effects depend on the mass of the gas in the disk.
Therefore, it is natural that the status of the ISM is related to the origin of the AVR. 
In other words, the AVR involves the historical evolution of the ISM in a galaxy.

\subsubsection{Dynamical Heating Rate}
\label{sec:HeatingRate}

In the previous section, we discussed the ZAVD. 
Here, we investigate the evolution of the dynamical heating rate (i.e. $d\sigma/dt$). 
Once these two factors are understood, 
we can use them to predict the entire evolution of the AVR.

\begin{figure}
\begin{center}
\includegraphics[width=0.48\textwidth]{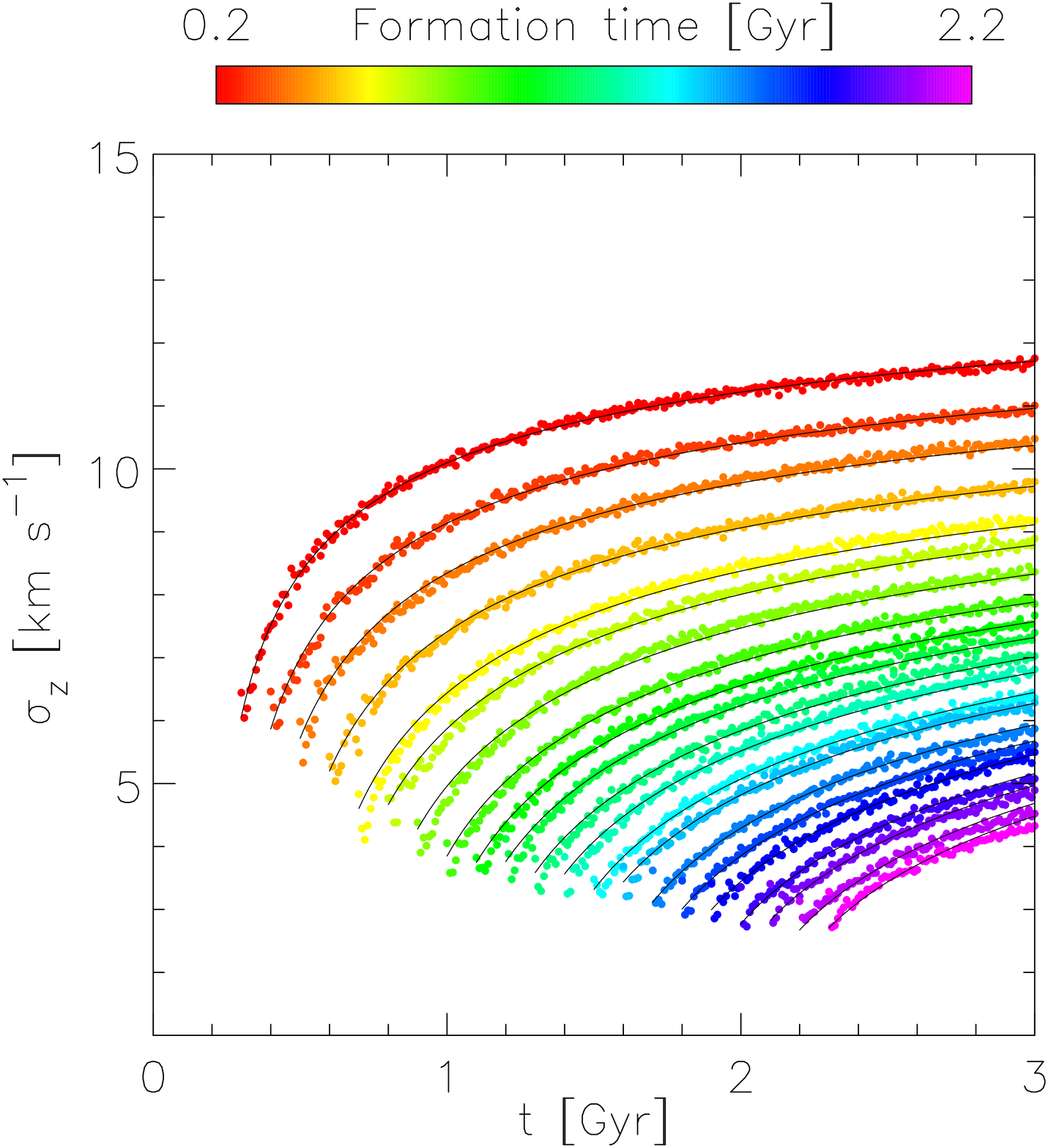}
\end{center}
\caption{
Vertical velocity dispersion as a function of time.  
Twenty-one groups have been selected. 
Each sequence represents the evolution of each group.
Solid curves are results of the fitting (see Eq. (\ref{eq:fit_sigma})).
}
\label{fig:t_sigma}
\end{figure}

Figure \ref{fig:t_sigma} shows the evolutions of vertical velocity dispersions,
$\sigma_{\rm z}$, as a function of simulation time.  
Each sequence is characterized by two evolution phases: 
the rapidly increasing phase in the early stage and 
the asymptotically flattening phase in the late stage.
Later time or higher velocity dispersion 
cause the heating rate to be lower.

\begin{figure*}
\begin{center}
\includegraphics[width=0.96\textwidth]{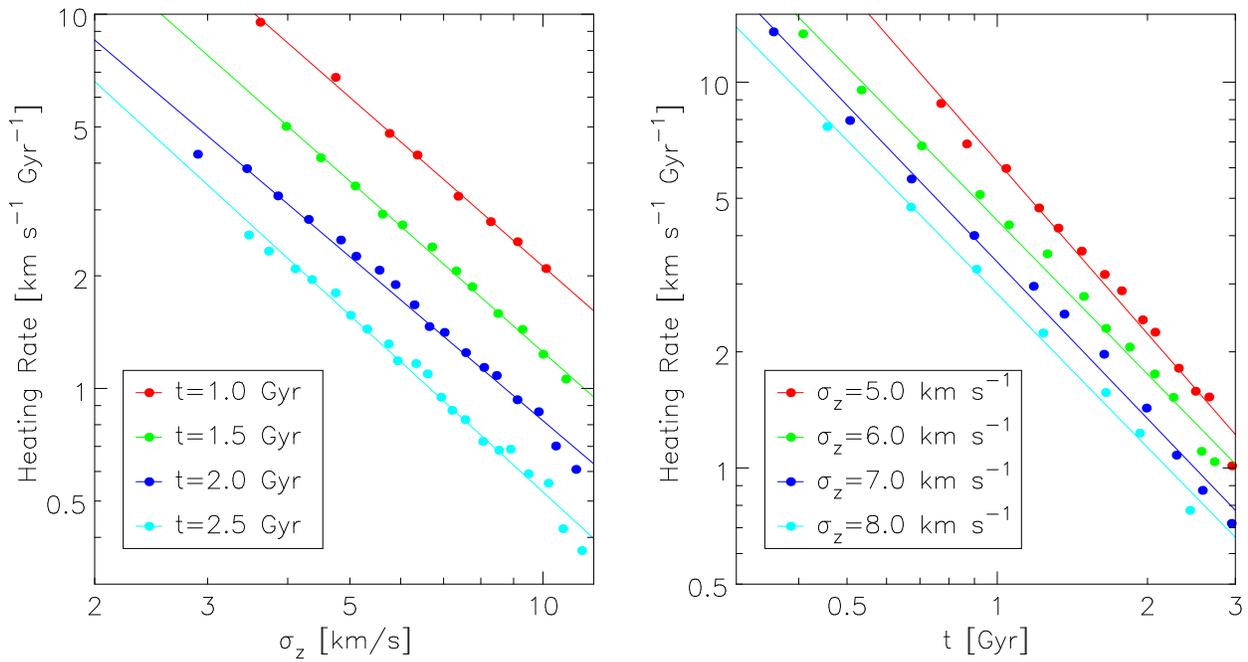}
\end{center}
\caption{
Heating rates as a function of $\sigma_z$ (left) and as a function of time $t$.
Four representative epochs ($t=1.0$, $1.5$, $2.0$ and $2.5~\rm{Gyr}$) and 
four representative values of $\sigma_z$ ($\sigma_z=5.0$, $6.0$,
$7.0$ and $8.0~{\rm km~s^{-1}}$) are employed.
The solid lines represent the results of fitting with power-law functions.
}
\label{fig:heating_rate}
\end{figure*}

Figure \ref{fig:heating_rate} shows the vertical heating rate (${\rm d}\sigma_z/{\rm d}t$) as
a function of $t$ and $\sigma_{\rm z}$.  We chose four representative epochs and
$\sigma_z$. Solid lines are the fitting result with power-law functions of
$\sigma_z$ and $t$. We can see that the heating rate obtained by our simulation
is well fitted by power-law functions.

Using the fitting results shown in figure \ref{fig:heating_rate}, we can
describe the heating rate as a function of $t$ and $\sigma_z$:
\begin{equation} 
  \frac{d\sigma_z}{dt} \propto t^{-\alpha} \sigma_z^{-\beta}.
  \label{eq:hr} 
\end{equation}
We apply this equation to sequences in figure \ref{fig:t_sigma} and then we obtain
the concrete values of $\alpha$ and $\beta$. With these values, we have
\begin{equation} 
  \frac{d\sigma_z}{dt} \propto t^{-1.3} \sigma_z^{-1.5}. 
  \label{eq:fit_heat} 
\end{equation}
By integrating this equation, we finally have 
\begin{equation} 
  \sigma_z = 11.8 \left[ A-\left( \frac{t}{1~{\rm Gyr}} \right) ^{-0.33} \right]^{0.40}~{\rm km~s^{-1}},
  \label{eq:fit_sigma} 
\end{equation}
where $A$ is a constant of an integral.
The different sequence found in figure \ref{fig:t_sigma} can be expressed by
choosing a different value of $A$.

\begin{figure}
\begin{center}
\includegraphics[width=0.48\textwidth]{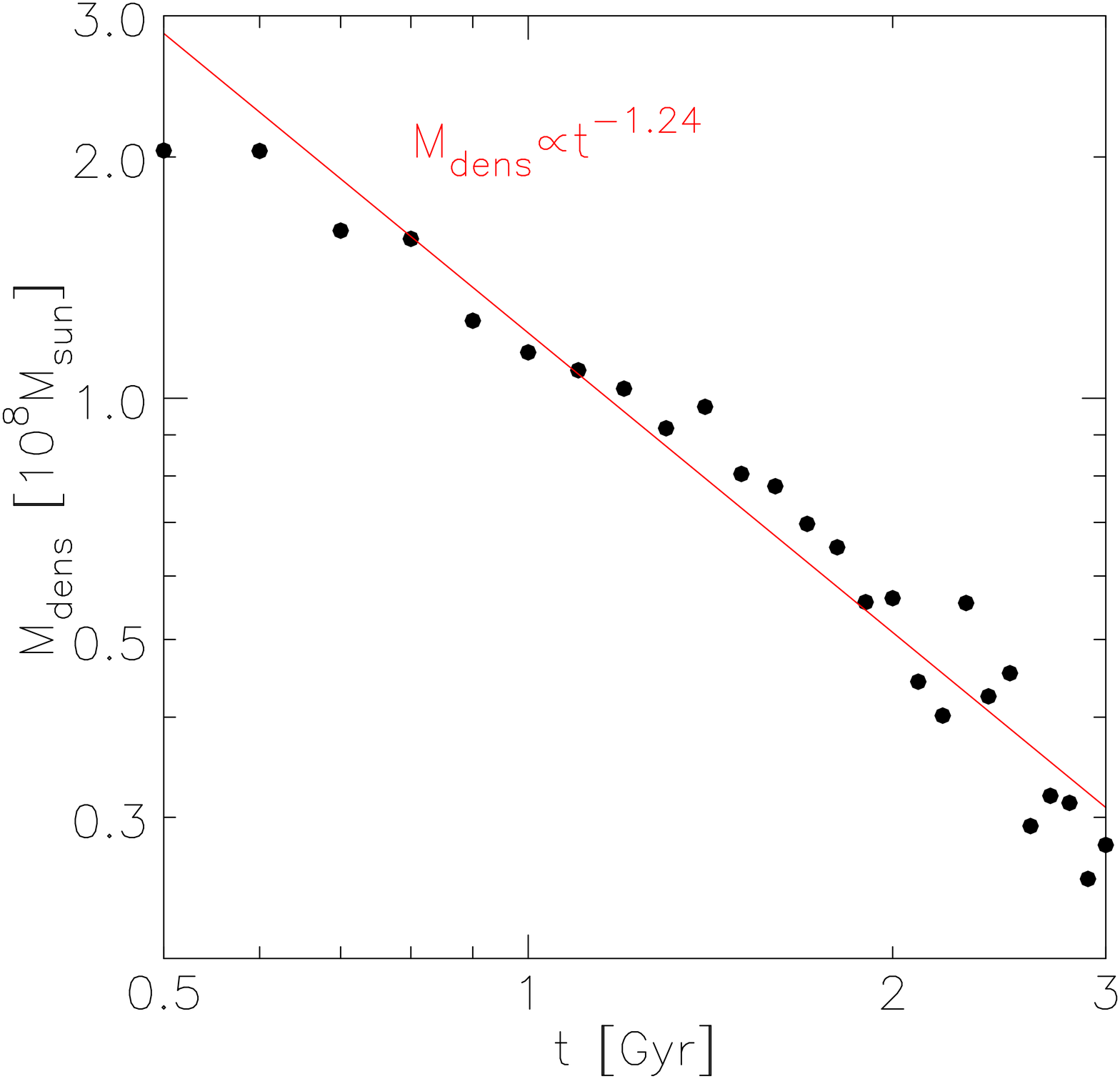}
\end{center}
\caption{
Time evolution of gas mass that is much denser than the star formation
threshold density.  The solid line indicates the result of the fitting by a
power-law function and its exponent is $-1.24$.
}
\label{fig:cold_mass}
\end{figure}

In figure \ref{fig:ssG}, the dense gas ($n_{\rm H} > 100~{\rm cm^{-3}}$) has clumpy structures, 
and is expected to play the role of a heating source.
This clumpy and dense gas is decreasing with simulation time.
We suspect that the decreasing heating rate with increasing simulation time 
may result from the decrease in the dense gas.
Hence, we show the time evolution of the dense gas mass $M_{\rm dens}$ in figure
\ref{fig:cold_mass}. Here, we define the gas above the star formation threshold
as the {\it dense gas}. We see that the mass of the dense gas
monotonically decreases and it can be fitted by a simple power-law function with
the power-law index of $-1.24$.  Interestingly, the power-law index is
consistent with that for the simulation time found in Eq. (\ref{eq:fit_heat}).
This implies that the time evolution of the heating rate in our simulation is
controlled by the total mass of the dense gas.

Our result can be reduced to that obtained in the previous study 
where the contribution of gas is unchanged \citep{SpitzerSchwarzschild1951,
SpitzerSchwarzschild1953,KokuboIda1992,HanninenFlynn2002}.  
Consider the case in which the dense gas mass does not evolve; 
this can be described by letting $\alpha=0$ in Eq. (\ref{eq:hr}),
and then we obtain 
\begin{equation} 
  \sigma_z \propto \left( A-t \right)^{1/(\beta+1)} .
  \label{eq:vdz2} 
\end{equation}
This is consistent with the time evolution of velocity dispersion shown by a
previous work (e.g. \cite{HanninenFlynn2002}), 
where the models treated the GMCs as a time-independent potential.

\section{Discussion and Summary}
\label{sec:Summary}

\begin{figure*}
\begin{center}
\includegraphics[width=0.96\textwidth]{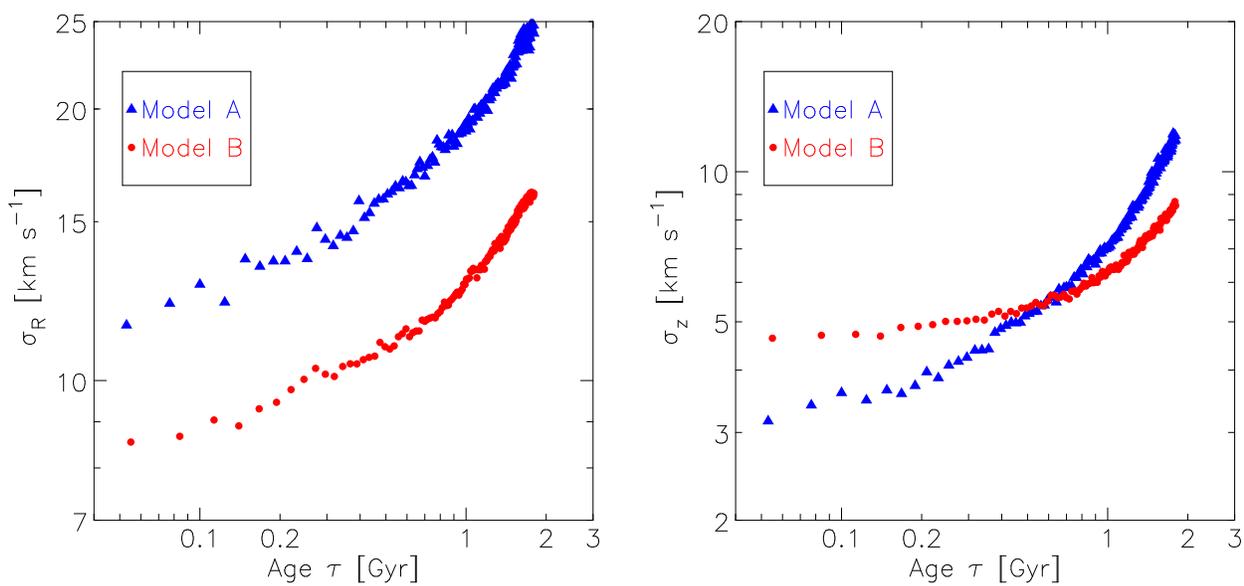}
\end{center}
\caption{Radial and vertical AVR of two simulations.
Blue and red points show the AVR of Models A and B, respectively.
}
\label{fig:avr_reso}
\end{figure*}

\begin{figure*}
\begin{center}
\includegraphics[width=0.96\textwidth]{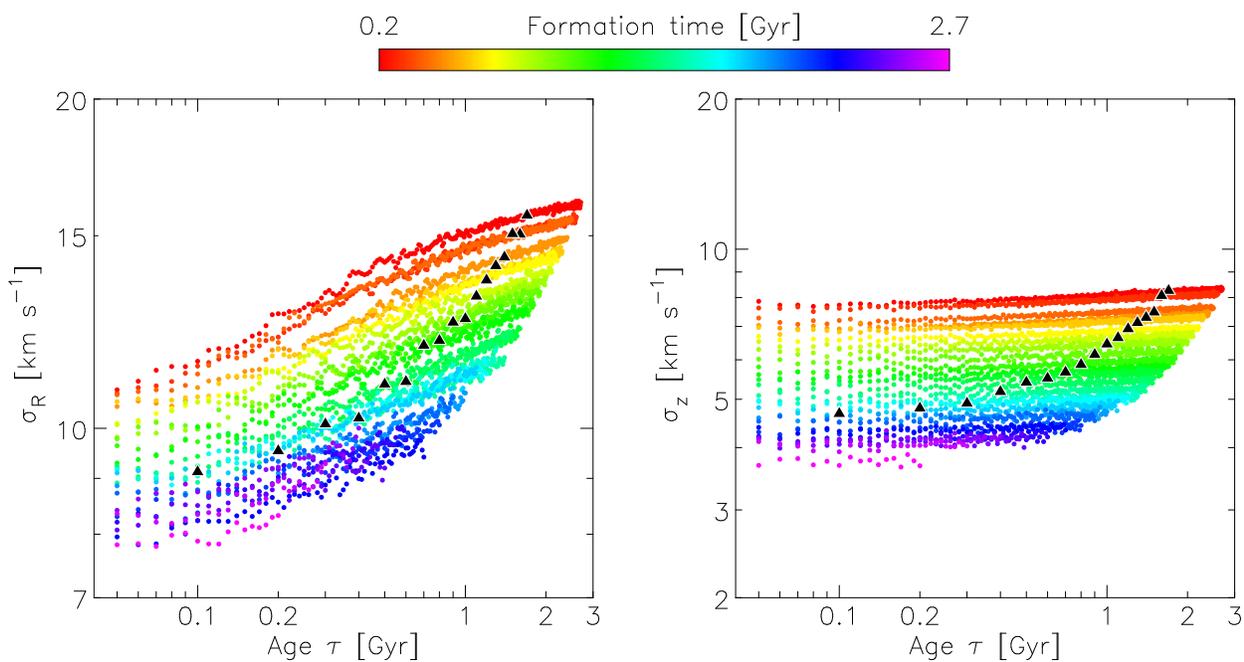}
\end{center}
\caption{Same as Figure \ref{fig:AVR:Evolution2}, but about a Model B whose
gravitational softening length and star formation density threshold are $100 \rm{pc}$ and $1 \rm{cm^{-3}}$ respectively.
}
\label{fig:origin_reso}
\end{figure*}

Our simulation shows that the AVR is not the evolution track of stars on the
age--velocity dispersion plane.  
This result means that the AVR can be reproduced
even if the heating rate is smaller than the slope of the AVR.  
The time evolution of the stellar velocity dispersion depends upon the combination of 
the initial velocity dispersion and the heating rate.  
In particular, the AVR is strongly affected by the evolution of the ZAVD.

The importance of the ZAVD was pointed out by \citet{Wielen1977};
however they did not attempt to elucidate the details because of the complicated nature of gas dynamics.
Recent numerical simulations of galaxy formation also demonstrate that 
the ZAVD has a crucial role in the establishment of the AVR \citep{Bird+2013, Grand+2016}.
The observations of turbulent gas disk at high redshift \citep{ForsterSchreiber+2009}
support the presence of the evolution of ZAVD.

The other key process for establishing the AVR is the secular heating  due to dense gas. 
This is consistent with \citet{Aumer+2016b}, who shows that the evolution of heating history shapes an AVR.
Our simulation can successfully derive the contribution of this process. 
We point out that the sufficient spatial resolution and 
realistic star-formation thresholds are crucial to describe the evolution of the AVR 
via dense gas scattering. 

\begin{table*}[htb]
  \caption{Parameters of simulation models.}
  \begin{center}
  \begin{threeparttable}
  \begin{tabular}{l|ccc} 
    \hline
    Models   & $\epsilon$ \tnote{1} [pc]   & $n_{\rm th}$ \tnote{2} [${\rm cm^{-3}}$]   & $N_{\rm ini}$ \tnote{3}  \\ 
    \hline\hline 
    Model A  & 10                          & 100                                    & $10^6$   \\
    Model B  & 100                         & 1                                      & $10^6$   \\
    \hline
  \end{tabular}
    \begin{tablenotes}\footnotesize
      \item[1] Softening length
      \item[2] Star formation density threshold
      \item[3] Initial particles
    \end{tablenotes}
  \end{threeparttable}
  \end{center}
  \label{tab:models}
\end{table*}

Figure \ref{fig:avr_reso} shows the radial and vertical AVRs of two simulations: 
one is obtained by our simulations with the fiducial parameters (Model A) 
and the other employs the larger gravitational softening length ($100~{\rm pc}$) 
and lower star-formation thresholds ($1~{\rm cm^{-3}}$) that are widely used in  
cosmological simulations of galaxy formation (Model B).
Table \ref{tab:models} shows parameters of the two models.
It is apparent that the AVRs with the low thresholds are quite different from those
obtained by our fiducial model; for instance, the ZAVDs for both radial and vertical
directions are different. The low threshold density for star formation decreases the radial ZAVD 
while it increases the vertical ZAVD. 
The former is due to the lack of high-density structures, and the latter is 
due to the relatively thick star-forming region in the gas disk (see \cite{Saitoh+2008}). 
Interestingly, the heating rate in the radial direction is almost identical to
that of the reference simulation while that in the vertical direction is quite
different from that of the reference simulation.  
The reason for the discrepancy in the vertical direction is the larger ZAVD.

Figure \ref{fig:origin_reso} shows the evolution tracks of different groups
for the low-resolution model. In particular, the evolution tracks of the vertical direction 
are quite different from those in the reference model (see figure \ref{fig:AVR:Evolution2}).
The adoption of star-formation thresholds, which are justified by high mass resolution,
is critical to any discussion of the formation and evolution of the AVR.
We will show the results of the systematic survey of the numerical effects on
the AVR in a forthcoming paper.

So far, we have worked with a very simplified model; for instance, 
there is no gas accretion, and the fixed halo and stellar potentials are used.  
We need to improve our model and make it more realistic so that 
we can discuss the long-time evolution of the AVR in more detail 
and can measure the contribution of spiral arms and the bar.
The contribution of the spiral arms to the AVR may turn out to be limited.
Previous studies have suggested that transient, recurrent (so-called `dynamic') spiral arms 
lead to radial migration and it does not increase the radial velocity dispersion significantly 
\citep{SellwoodBinney2002,Grand+2012a,Grand+2012b,Roskar+2012,Baba+2013}.

As a next step, we will introduce the effects of gas accretion to our model. 
By investigating the relation between the accretion history and the AVR, 
it might be possible to impose constraints on the Milky Way-like galaxy formation.

The GAIA era is approaching. The GAIA will provide information about the proper motions 
of about one billion stars \citep{Perryman+2001}. 
By combining it with other observations, such as the RAVE \citep{Steinmetz+2006},
APOGEE \citep{Majewski+2016}, SEGUE \citep{Yanny+2009}, and Gaia-ESO survey \citep{Gilmore+2012},
we will be able to depict the fine structure of the Milky Way galaxy.  
Our model can be tested by the results of this new generation of observations. 

\bigskip

We thank the referee for his/her constructive comments which helped improve the manuscript.
Numerical computations in this paper were performed on Cray XC30 at the Center for
Computational Astrophysics, National Astronomical Observatory of Japan.  TRS is
supported by a Grant-in-Aid for Scientific Research (26707007) of Japan Society
for the Promotion of Science.
JB was supported by HPCI Strategic Program Field 5 `The origin of matter and the universe'
and JSPS Grant-in-Aid for Young Scientists (B) Grant Number 26800099.


\end{document}